**The Bell inequality, inviolable by data used consistently with its derivation, leads to quantum correlations that satisfy it, and probabilities that satisfy the Wigner inequality.**

Louis Sica[1,2]

[1]Institute for Quantum Studies, Chapman University, Orange, CA & Burtonsville, MD, 20866, USA
[2]Inspire Institute, Inc., Alexandria, VA 22303, USA
Email: lousica@jhu.edu

It is not generally known, that the inequality that Bell derived using three random variables must be identically satisfied by any three corresponding data sets of ±1'$s$ that are writable on paper. This surprising fact is not immediately obvious from Bell's inequality derivation based on causal random variables, but follows immediately if the same mathematical operations are applied to finite data sets. For laboratory data, the inequality is identically satisfied as a fact of pure algebra, and its satisfaction is independent of whether the processes generating the data are local, non-local, deterministic, random, or nonsensical. It follows that if predicted correlations violate the inequality, they represent no three cross-correlated data sets that experimentally exist or can be generated from valid probability models. Reported data that violate the inequality consist of probabilistically independent data-pairs and are thus inconsistent with inequality derivation. In the case of random variables as Bell assumed, the correlations in the inequality may be expressed in terms of the probabilities that give rise to them. A new inequality results, the Wigner inequality that must be satisfied by quantum mechanical probabilities in the case of Bell experiments. If that were not the case, predicted quantum probabilities and correlations would be inconsistent with basic algebra.

**1. Introduction**

Bell constructed an inequality from the computation of three cross-correlations of three simultaneously existing random variables [1], and applied it to an experiment that produced only two variables per random realization. The properties of the third variable result from its mathematical definition. The purpose of the third variable was to provide insight into the random quantum mechanical state of entanglement, and evidence as to whether or not predicted correlations could be accounted for by random but causal variables. The mathematical construction uses correlations of readouts equaling ±1 (as detector clicks are labeled) that are functions of variables that might be reasonably assumed to be randomly occurring initial conditions. (The experimental schematic of classic Bell experiments is shown in Fig. 1.) To analyze physical aspects of the situation, Bell postulated one measurement on the left side of the apparatus and two measurements at mutually exclusive settings on the right side [2]. However, only one particle is produced on each side, and each is destroyed by observation. A key question immediately arises: how is the inequality to be applied to the experimental situation that it has been designed to address?

There are only two ways known to the author to increase the number of experimental variables from two to three to enable application of the Bell inequality. One way is to add a polarizer and two detectors in each polarizer output on one side in Fig. 1. A final detection then reveals which of the four possible paths on that side of the apparatus was taken using an analysis known as retrodiction. In this case the probability of the last path segment depends conditionally on the random selection of the preceding path segment, and the resulting three correlations are unequal [3]. This is described by the quantum principle that two successive polarization measurements on a photon are non-commutative.

Bell specifically rejected the above choice of realization of variables for application of his inequality [2], and stated that the second random readout at a different polarizer setting was to be realized as the readout that would have occurred in the same random trial if that alternate setting had been used in place of the first setting: "the 'spin' of each particle is measured once only." *Thus, the construction of the three random-variable Bell inequality, based on identical hidden variables for each of three observations, one of which is unperformed, is inapplicable to experimental data, since one cannot undo* *results at one instrument setting in a random trial to obtain results that would have occurred at a different instrument setting.*

However, it will be shown in Sec. 2 that the inequality that Bell derived holds for physical data under far more general conditions than Bell originally realized [4,5], and in a form leading to its satisfaction under realizable experimental conditions. (This is also true for the four variable version of the inequality.) Given that three data sets, random or deterministic, have been experimentally obtained so as to be writeable on paper, the same algebraic steps that Bell used to derive his statistical inequality may be applied to them. The Bell inequality again results, but is now identically satisfied by cross-correlations of three finite sets of data with all items equal to $\pm 1$. Thus, if applied to actual physical data for three variables, the Bell inequality is not statistical or probabilistic, but is an identity-inequality of algebra that may then be applied to random or deterministic situations. (This same generally unrecognized fact holds for the four variable inequality.) The correlation estimates that result when applying the inequality to three data sets of $\pm 1's$ may now statistically converge to three different functional forms. It is these functional forms that are experimentally verifiable, and not whether the identity-inequality is satisfied, since it must be satisfied by any three data sets. Finally, since the data are random in the case of Bell experiments, the resulting correlations may be expressed in terms of the probabilities that give rise to them, yielding an inequality in probabilities, the Wigner inequality, that must also now be satisfied. This inequality is often presented as logically independent of the Bell inequality but this cannot be so when both are applied to the same physical situation.

Given that the general form of the Bell inequality is purely algebraic involving three physical data sets, and that it must be identically satisfied by such data sets, the remaining problem is to create a procedure for acquiring appropriate data from Bell experiments that create data in pairs. Since the inequality under Bell's assumptions cannot be experimentally realized, a procedure must be found that provides data at two mutually exclusive settings for a B-side particle in Fig. 1 corresponding to two fixed outputs on the A-side. Such data may be acquired using two experimental runs [6] at fixed setting *a* to allow two *B*-side data sets to be obtained for each output on the *A*-side. The two experimental trials at mutually exclusive settings of *B* are now probabilistically independent except for their conditional dependence on a common output at *A*. (This situation is known as "conditional independence" in mathematics [7].) The two cross-correlated observations on the *B*-side now have a different form than their correlation with the output at *A*. Three data sets result that correspond to the three variables in Bell's derivation. The variables are correlated precisely as Bell's original variables, but the hidden variables for each correlation (if assumed to exist) must now be different since they occur in different trials.

To understand key aspects of this more simply, consider flipping a loaded coin with load 1. After a given flip, one could ask the equivalent of Bell's question: suppose the load had been load 2 on that same flip, what would the outcome have been? This is clearly an unanswerable question in the case of causal random variables as assumed in Bell's representation, unless they are all under total control and can be fully analyzed, as is unrealistic in random experiments. The common way of addressing this situation is to flip the coin a large number of times with each loading and determine the probabilities of heads and tails for the two cases. It should be observed that in the case of continuous variables leading to a pair of discrete events, a range of variable values leads to the same outcome, but a change in any decimal place in a variable at the boundary between outcomes could lead to a change in outcomes. Thus, in a world of finite precision instruments, causality does not necessarily imply predictability.

The incorrect assumption by Bell, in the absence of actual calculation, that the third correlation in his statistical inequality had the same functional form as the first two in either theoretical or experimental realization has no logical basis, and has been a fundamental source of confusion for more than fifty years. (The same problems discussed above affect the four variable inequality but are necessarily more complex to deal with.) This assumption, inconsistent with the derivation of the inequality, has led experimentalists to compute correlations in individual pairs. In the case of a wide-sense stationary process [8] for which any number of variables may be measured and all pairs have a correlation of the same form, the Bell inequality would not be violated. This special form of stochastic process, widely used in optics as an approximation, does not hold in the Bell experiment case.

## 2. The constant in Bell's inequality in three variables results from the use of three variables

The facts underlying the above discussion will now be exhibited in detail beginning with a review of Bell's derivation of the inequality. Bell considered the output of an ideal source of entangled spins. In experiments, photons have been used in place of spins due to the development of very efficient Bell-correlated photon sources [9]. From the experimental setups (see Fig. 1) two photons at a time emerge traveling in different directions. Given Bell's rejection of added polarizers and detectors that would allow obtaining three measurements at once, only two polarization paths at a time through the apparatus may be obtained.

Measurements on the A-side of the apparatus at angular setting $a$ are represented in Bell's notation by the function $A(a,\lambda)$, and on the B-side at angular setting $b$ by $B(b,\lambda)$. The variables $\lambda$ are random variables with a probability density $\rho(\lambda)$ assumed to determine the results of measurements represented by functions $A$ and $B$, postulated to be deterministic. The random results of measurements might then be interpreted causally as due to uncontrolled initial conditions sampled randomly. The measurement variables $A$ and $B$ have values equal to ±1 with the additional requirement $B(a,\lambda) = -A(a,\lambda) = \pm 1$, so that required measurement outcomes resulting from entanglement are fulfilled [1].

To obtain additional conditions on the correlations of readouts, Bell assumed an output at one setting on the $A$ side of the apparatus and corresponding outputs at two alternative settings on the $B$-side using the same hidden variable values for each. Since only one actual measurement is to be performed per side, given Bell's prescription, two experimental realizations of photons must ultimately be used to obtain such data but that would imply different values for the assumed hidden variables in the two realizations. Thus, correlations satisfying Bell's conditions are physically unrealizable. A more general derivation of the inequality below that is not based on a specific representation of hidden variables eliminates this difficulty.) In Bell's derivation the correlation of an A-side with a B-side measurement is given by

$$C(a,b) = \int A(a,\lambda) B(b,\lambda) \rho(\lambda) d\lambda . \qquad (2.1)$$

The difference between two such correlations is

$$|C(a,b) - C(a,b')| = \left|\int (A(a,\lambda) B(b,\lambda) - A(a,\lambda) B(b',\lambda)) \rho(\lambda) d\lambda\right|$$

$$\leq \int |A(a,\lambda) B(b,\lambda)| |(1 - A(a,\lambda) B(b,\lambda) A(a,\lambda) B(b',\lambda))| \rho(\lambda) d\lambda =$$

$$\leq \int |(1 - B(b,\lambda) B(b',\lambda))| \rho(\lambda) d\lambda = 1 - C(b,b'), \qquad (2.2)$$

where $b'$ is used for the second variable consistent with Bell's explicit prescription that two alternative measurements are theoretically considered as occurring on one side of the inequality [2]. The final correlation follows from the fact that

$$A(a,\lambda) B(b,\lambda) A(a,\lambda) B(b',\lambda) = B(b,\lambda) B(b',\lambda) , \qquad (2.3)$$

since $A(a,l)^2 = 1$. (A similar result holds for the fourth correlation in the four variables inequality [10].) Thus, the third correlation is obtained in a different way than the first two correlations $C(a,b)$ and $C(a,b')$. This widely unrecognized but crucial fact has recently also been noted by Hess [11] in a review of Bell's theorem and related topics. Hess found that it was also established in a random variables context by Vorobev[12].

While in the last line of (2.2) $C(b,b')$ superficially appears to be the result of an independent measurement process and to have the same dependence on hidden variables as do the other correlations, its computation in fact reuses the data from the previous correlations, data pair by data pair. Thus, it would not be surprising if the correlation $C(b,b')$ had a different functional form from the others even under Bell's theoretical construction. If a result at $b$ is obtained and continuous causal variables are involved in its outcome, then given that the ranges of those variables are now restricted due to causing that particular outcome, possible outcomes at $b'$ are now affected. Thus, a result at $b'$ would be conditionally dependent in probability on an outcome at $b$ as well as $a$, and this does not appear to have been considered, given Bell's notation. While such results may be computed, based on assumed mathematical models of hidden variables, they are still un-measurable unless the variables can be physically controlled with mathematical precision. The author proposes an experimental resolution of this problem, based on a more general form of Bell's inequality now to be given.

The possible laboratory realization of (2.2) is conceptually facilitated if one considers that to compute Bell's desired correlation functions, three data sets must somehow be acquired and written on paper, one data set for each of Bell's variables. (Outputs occur at settings *a*, *b*, and *b'*, with that at *b'* ultimately recorded on the opposite side at *a'=b'*, thus resulting in a minus sign before the final correlation in Bell's version.) Note that while correlations can be directly measured in some optical experiments, they are not measured in Bell experiments. Only detector clicks are observed from which correlations are later computed. Assume that the data sets, random or deterministic, are labeled by instrument settings $a$, $b$, and $b'$. The corresponding data set items are denoted by subscripted variables $a_i$, $b_i$, and $b'_i$ with $N$ items in each set. Each datum equals $\pm 1$. The data actualization of (2.2) begins with

$$a_i b_i - a_i b'_i = a_i b_i (1 - a_i b_i a_i b'_i) . \qquad (2.4)$$

where the unusual factorization holds only for the specific values of the variables considered, each equal to $\pm 1$. Summing (2.4) over *i* from 1 to *N* and taking absolute values of both sides yields

$$\left| \frac{1}{N} \sum_{i=1}^{N} (a_i b_i - a_i b'_i) \right| = \left| \frac{1}{N} \sum_{i=1}^{N} a_i b_i (1 - a_i b_i a_i b'_i) \right| \leq \frac{1}{N} \sum_{i=1}^{N} |(1 - a_i b_i a_i b'_i)|$$

$$\frac{1}{N} \sum_{i=1}^{N} a_i b_i - \frac{1}{N} \sum_{i=1}^{N} a_i b'_i \leq 1 - \frac{1}{N} \sum_{i=1}^{N} a_i b_i a_i b'_i = 1 - \frac{1}{N} \sum_{i=1}^{N} b_i b'_i , \qquad (2.5)$$

the same result as (2.2) but now seen to hold for the resulting correlation estimates using the actual finite data sets independent of any hidden variables. Note that (2.5) does not depend on whether the processes resulting in the data are local, nonlocal, random or deterministic: no assumptions have been made regarding the physical character of the data other than that each item equals $\pm 1$, and all three data sets are available to compute the cross-correlations. Thus, result (2.5) holds even if there is nonlocal "pickup" between detectors so that the *B* setting affects *A*, etc. Note that the correlation estimates may converge to three different functional forms and satisfy (2.5). Thus, the resulting inequality is identically satisfied if Bell's mathematical steps are applied to actual physical data. Note that just as in (2.3), data pair by data pair, $a_i b_i a_i b'_i = b_i b'_i$ so that the final correlation is determined by reused data from the first two correlations. *The third correlation does not result from an independent data acquisition process.*

Under casual examination, the last line of (2.2) might suggest that the variables used to compute the three correlations may be measured and correlated in three separate variable pairs, since their dependence on hidden variables is the same in the final correlation integrals. However, the physical procedure used to obtain the data indicates a different conclusion. The constant in the inequality arises from the fact that the same value of $a_i$ multiples $b_i$, and $b'_i$, the same value of $b_i$ multiples $a_i$ and $b'_i$, the same value of $b'_i$ multiplies $a_i$ and $b_i$. Thus, the three variable values must all be available at the time the cross-correlations are calculated. (A similar algebraic condition holds in the case of the four variable inequality that is also easily shown to be an inequality-identity.) While the variables are symmetrically treated in an algebraic sense, their condition of physical acquisition is quite asymmetric and, as will be seen, results in different functional forms for the correlations.

Failure to consider the affects of the measurement procedure on the algebraic relations between variables has led to the processing of measurements for correlations in individual pairs [13], and thus mathematical inconsistency [14,15] with inequality derivation. Given the inconvenience of applying a three variable inequality to a random process producing two variables per realization, the mathematical conditions under which the inequality holds have in practice been neglected. Six random variables have been acquired in three random trials to obtain three independent correlations, whereas the three variable inequality, as demonstrated above in two derivations, depends on the use of three cross-correlated random variables to obtain three correlations. Further, since the third correlation depends on the first two measurement procedures and reuses their previously acquired data, it is computed under a different data acquisition procedure than the first two, and is therefore expected to have a different form.

**3. How Can Bell's Inequality Be Applied to Experiments?**.

Bell constructed (2.2) to apply to results of experiments schematized in Fig. 1. However, the correlations $C(ab)$ and $C(ab')$ require data at mutually exclusive settings $b$ and $b'$ of a polarization beam splitter. Two separate experiments are required to obtain such data. To apply (2.2) requires that the setting $a$ be the same on each trial, and that outputs at $b$ and $b'$ be recorded for each output value at setting $a$. Conditions for applicability of the outcomes of (2.2) and (2.5) are then satisfied, and one finds that the resulting correlation for $C(bb')$ has a different form from that of $C(ab)$ and $C(ab')$, a fact not recognized by Bell, or later by experimentalists.

The correlation $C(ab)$ is easily derived using well known quantum mechanical probabilities resulting from entanglement [16]. The subscripted pluses and minuses of the probabilities below indicate $\pm 1$ outputs at instrument settings $a$ and $b$, respectively:

$$P_{++}(a,b) = P_{--}(a,b) = \tfrac{1}{2}\sin^2\frac{b-a}{2}; \quad P_{+-}(a,b) = P_{-+}(a,b) = \tfrac{1}{2}\cos^2\frac{b-a}{2}. \tag{3.1}$$

From these probabilities

$$C(a,b) = [(+1)(+1)+(-1)(-1)]\tfrac{1}{2}\sin^2\frac{b-a}{2} + [(+1)(-1)+(-1)(+1)]\tfrac{1}{2}\cos^2\frac{b-a}{2}$$

$$= -\left(\cos^2\frac{b-a}{2} - \sin^2\frac{b-a}{2}\right) = -\cos(b-a), \tag{3.2}$$

where $C(ab')$ follows immediately by replacing $b$ with $b'$.

Computation of $C(bb')$ requires the conditional probabilities $P(b|a)$ and $P(b'|a)$ obtainable from joint probabilities (3.1) already written in terms of conditional probabilities resulting from entanglement since $P_+(a) = P_-(a) = 1/2$. The data from separate runs at independent settings $b$ and $b'$ are correlated and conditionally dependent on each of the outputs of $\pm 1$ at $a$. Thus, for the +1 case,

$$C(bb'|a,1) = \sum_{b_i b_i'} b_i b_i' P(b_i, b_i'|a,1) = \sum_{b_i b_i'} b_i b_i' P(b_i|a,1) P(b_i'|a,1) \tag{3.3}$$

where the subscripted variables equal plus or minus 1, and in the notation of (3.3), $a$ is indicated as having output +1. Since the probabilities describe outputs in independent trials and at different variable settings, the probability $P(b_i, b_i'|a,1)$ factors [7]. Thus,

$$C(bb'|a,1) = \sum_{b_i} b_i P(b_i|a,1) \sum_{b_i'} b_i' P(b_i'|a,1) = \overline{b}(a,1)\overline{b}'(a,1). \tag{3.4}$$

Using the conditional probabilities obtained from (3.1), one obtains

$$\overline{b}(a,1) = 1P(b,1|a,1) - 1P(b,-1|a,1) = \sin^2(1/2)(b-a) - \cos^2(1/2)(b-a) = -\cos(b-a),$$
$$\overline{b}'(a,1) = 1P(b',1|a,1) - 1P(b',-1|a,1) = \sin^2(1/2)(b'-a) - \cos^2(1/2)(b'-a) = -\cos(b'-a), \tag{3.5}$$

and from (3.4),

$$C(bb'|a,1) = \cos(b-a)\cos(b'-a). \tag{3.6}$$

The same result is obtained for $C(bb'|a,-1)$ so that [17]

$$C(bb') = \sum_i C(bb'|a_i)P(a_i) = (1/2)C(bb'|a,1) + (1/2)C(bb'|a,-1) = \cos(b-a)\cos(b'-a). \tag{3.7}$$

Inserting (3.7) and Bell correlations into (2.2) yields:

$$0 \leq \cos(b-a) - \cos(b'-a) + 1 - \cos(b-a)\cos(b'-a)$$
$$0 \leq \cos(b-a)2\left(\frac{1}{2} - \frac{1}{2}\cos(b'-a)\right) + 2\left(\frac{1}{2} - \frac{1}{2}\cos(b'-a)\right) \quad (3.8)$$
$$0 \leq 4\left[\sin^2(\frac{b'-a}{2})\right]\left[\cos^2(\frac{b-a}{2})\right].$$

Thus, Bell's inequality is satisfied.

## 4. The Wigner inequality in probabilities results from the Bell inequality when both apply to the same physical situation

### 4.1 Bell variable re-definition

Two of the three variables in (2.4) have been given labels $b_i$ and $b'_i$ since as Bell indicated, they represent values of alternative measurements on the right-hand $B$-side of a Bell experiment apparatus (Figure 1). The final term in (2.5) may be written

$$\frac{\sum_i^N b_i(-1) \cdot (-1)b'_i}{N} = -\frac{\sum_i^N b_i(-1)b'_i}{N} = -\frac{\sum_i^N b_i a'_i}{N} \quad (4.1)$$

where $a_I' = -b'_i$, for detector settings $a'$ and $b'$ that are equal but on opposite sides of the apparatus. (Given the properties of entanglement, measurements at the same angular settings on opposite sides of a Bell apparatus have opposite signs, i.e., $+1 \Rightarrow -1$ and $-1 \Rightarrow +1$.) Thus, using Bell measurements in (2.5) with the $b'$ variable in the right-most term replaced by an equal setting on the opposite side, one has

$$\left|\frac{\sum_i^N a_i b_i}{N} - \frac{\sum_i^N a_i b'_i}{N}\right| \leq 1 + \frac{\sum_i^N b_i a'_i}{N} \quad (4.2)$$

and

$$|C(a,b) - C(a,b')| \leq 1 + C_3(b,a') \quad (4.3)$$

assuming convergence of the correlation estimates as $N \to \infty$.

### 4.2 Probabilities corresponding to the correlations

Since probabilities must exist that determine the correlations between each of the variable pairs in (4.3), the correlations may be written in terms of those probabilities. The notation to be used for the probabilities is $P_{xy}(a,b)$, where x and y equal +1 or -1 and indicate the outputs at instrument setting angles $a$ and $b$ respectively. In the quantum mechanical case after assuming that (perfectly) entangled particle pairs produce the measurements, the probabilities are symmetrical as in (3.1) so that

$$P_{++}(a,b) = P_{--}(a,b) \text{ and } P_{+-}(a,b) = P_{-+}(a,b) \quad (4.4a)$$

with normalization condition

$$P_{++}(a,b) + P_{--}(a,b) + P_{+-}(a,b) + P_{-+}(a,b) = 1. \quad (4.4b)$$

Then

$$P_{+-}(a,b) = 1/2 - P_{++}(a,b) \quad (4.4c)$$

so that

$$C(a,b) = 2P_{++}(a,b) - 2P_{+-}(a,b) = 4P_{++}(a,b) - 1. \quad (4.4d)$$

The result for $C(bb')$ is similar since it will immediately be shown that

$$P_{+-}(b,b') = P_{-+}(b,b') \text{ and } P_{++}(b,b') = P_{--}(b,b'). \tag{4.4e}$$

This follows from (3.1):

$$P_{++}(b,b') = P_{++}(b,b'|a,1)P(a,1) + P_{++}(b,b'|a,-1)P(a,-1) =$$
$$\frac{1}{2}\left[P_+(b|a,1)P_+(b'|a,1) + P_+(b|a,-1)P_+(b'|a,-1)\right] = \tag{4.5a}$$
$$\frac{1}{2}\left[\sin^2\frac{b-a}{2}\sin^2\frac{b'-a}{2} + \cos^2\frac{b-a}{2}\cos^2\frac{b'-a}{2}\right],$$

with an equal result for $P_{--}(b,b')$. Similarly, $P_{+-}(b,b')$ equals

$$P_{+-}(b,b') = \frac{1}{2}\left[P_+(b|a,1)P_-(b'|a,1) + P_+(b|a,-1)P_-(b'|a,-1)\right] =$$
$$\frac{1}{2}\left[\sin^2\frac{b-a}{2}\cos^2\frac{b'-a}{2} + \cos^2\frac{b-a}{2}\sin^2\frac{b'-a}{2}\right], \tag{4.5b}$$

with an equal result for $P_{-+}(b,b')$ leading to normalization $2P_{++}\{b,b') + 2P_{+-}(b,b') = 1$. Then as in (4.4d):

$$C(b,b') = 2P_{++}(b,b') - 2P_{+-}(b,b') = 4P_{++}(b,b') - 1 \tag{4.5c}$$

Inserting results (4.4d) and (4.5c) into the Bell inequality yields the Wigner inequality [18, 19]:

$$4P_{++}(a,b) - 1 - (4P_{++}(a,b') - 1) \le 1 - (4P_{++}(b,b'|a) - 1)$$
$$P_{++}(a,b) - P_{++}(a,b') \le 1/2 - P_{++}(b,b') = P_{+-}(b,b') \tag{4.5}$$
$$\le P_{++}(b,a').$$

The last step follows from the use of the variable $a'$ ($a' = b'$) on the opposite side of the apparatus from $b'$ to reverse the sign of the output due to entanglement, i.e., $P_{+-}(b,b') = P_{++}(b,a')$.

The Wigner inequality is usually derived from purely probability assumptions [18] based on entanglement that do not address the fact that the right-side probability is conditional on the left-hand side outcomes, given the conditions necessary to obtain two independent measurements on one side of the apparatus. The probability calculations to obtain the Wigner inequality ultimately result from the operations necessary to obtain a physical realization of Bell's variables.

## 5. Quantum mechanical probabilities satisfy the Wigner inequality

It now must be shown that the probabilities required by the experimental conditions for applicability of inequality (4.5) satisfy the inequality. From (3.1)

$$P_{++}(a,b) = (1/2)\sin^2\frac{1}{2}(a-b) \tag{5.1a}$$

and

$$P_{+-}(b,b') = P_{+-}(b,b'|a_+)P(a_+) + P_{+-}(b,b'|a_-)P(a_-) =$$
$$P_+(b|a_+)P_-(b'|a_+)P(a_+) + P_+(b|a_-)P_-(b'|a_-)P(a_-) \tag{5.1b}$$

From (4.5b)

$$P_{+-}(b,b') = (1/2)\left(\sin^2\frac{1}{2}(b-a)\cos^2\frac{1}{2}(b'-a) + \cos^2\frac{1}{2}(b-a)\sin^2\frac{1}{2}(b'-a)\right) \tag{5.2}$$

Note that $P_{+-}(b,b') = P_{-+}(b,b')$.

Inserting (5.1a) (and its equivalent for $b'$) together with (5.2) in (4.5) one obtains:

$$0 \leq -\sin^2\frac{(a-b)}{2} + \sin^2\frac{(a-b')}{2} + \sin^2\frac{(a-b)}{2}\cos^2\frac{(a-b')}{2} + \cos^2\frac{(a-b)}{2}\sin^2\frac{(a-b')}{2}$$
$$0 \leq 2\sin^2\frac{(a-b')}{2}\cos^2\frac{(a-b)}{2}.$$
(5.3)

Thus, the Wigner inequality is satisfied by quantum probabilities corresponding to a performable Bell experiment.

## 6. Computational counter-example

It has been shown above that the Bell inequality is satisfied by any three laboratory data sets without the assumptions of locality and hidden variables commonly thought to be necessary to construct it. It holds immediately for any three data sets that can be written on paper. That does not in itself, however, imply that independent random processes using common boundary conditions can produce Bell correlations. Nevertheless, a number of researchers have claimed achievement of such derivations [14]. (Bell correlations have also been computed based on very small information transfer from detector $A$ to detector $B$ [20].) The articles referred to are not physical derivations of the correlations, which are effectively limited by our unsettled understanding of photons and their relation to electromagnetic waves. However, derivations of Bell correlations based on more physical principles have also been given [21,22].

It thus seems fitting to conclude this logical assessment of the Bell and Wigner inequalities with a short algorithmic counter-example yielding a Bell cosine correlation. It will be developed from two independently constructed random variables and probabilities followed by a third random process that determines whether or not a two photon event has occurred. The latter corresponds to the spontaneous two-photon down-conversion process used in the source for classic Bell experiments. The model developed was suggested by an example in a Papoulis monograph [8] that begins with a non-stationary random process in two continuous angle variables and imposes a condition that makes the process stationary.

Assume two functions using arbitrarily chosen settings $\theta_1$ and $\theta_2$:

$$z_1 = a\cos\theta_1 + b\sin\theta_1$$
$$z_2 = -a\cos\theta_2 - b\sin\theta_2$$
(6.1)

with product

$$z_1 z_2 = -[a^2\cos\theta_1\cos\theta_2 + b^2\sin\theta_1\sin\theta_2 + ab(\cos_1\sin\theta_2 + \sin\theta_1\cos\theta_2)].$$
(6.2)

If $a$ and $b$ are independent random variables with the same probability density and zero mean, the average of (6.2) is

$$\overline{z_1 z_2} = -\overline{a^2}\cos\theta_1\cos\theta_2 - \overline{b^2}\sin\theta_1\sin\theta_2$$
$$= -\overline{a^2}\cos(\theta_1 - \theta_2),$$
(6.3)

where it will be assumed for use below that $\overline{a^2} = \overline{b^2} \ll 1$.

Expression (6.3) needs to be converted to one with variables equal to $\pm 1$ as necessary to associate photon detection counts with the computation:

$$\frac{z_1}{|z_1|}\frac{z_2}{|z_2|} = \frac{-(a\cos\theta_1 + b\sin\theta_1)(a\cos\theta_2 + b\sin\theta_2)}{[z_1 \| z_2|}$$
(6.4)

The factors being multiplied now equal $\pm 1$. But the average of (6.4) is not equal to that of (6.3). To accomplish this last step an uncertainty in count-pair production is introduced by assuming a Poisson process defined by the probability of a twin-photon event occurrence:. $P(1) = |z_2 \| z_2| \ll 1$. The probability of a zero or non-event is approximated by $P(0) = 1 - P(1)$. The average value of the product (6.4) is then its value averaged by its rate of occurrence ($P(a,b)$ yields the random behavior of $a$ and $b$ resulting in (6.3)):

$$\sum P(a,b)[P(1)1 + 0P(0)] \frac{\bar{z}_1}{|z_1|} \frac{\bar{z}_2}{|z_2|}$$

$$= \sum P(a,b)[|z_1||z_2|] \frac{\bar{z}_1}{|z_1|} \frac{\bar{z}_2}{|z_2|} = \overline{z_1 z_2} = -\overline{a^2} \cos(\theta_1 - \theta_2) \qquad (6.5)$$

In the above derivation, detector efficiency does not occur and values of $\theta_1$ and $\theta_2$ may be freely chosen. Thus, entanglement is not the sole generator of Bell cosine correlations. It should also be observed that Bell's formalism does not represent the above example since the variables $A(a,\lambda)$ and $B(b,\lambda)$ now result from a ratio of two functions controlled by separate, qualitatively different random processes. The overall process is more complex than that implied by the Bell notation. The physical processes considered in [21, 22] are even more complex.

## 7. Conclusion

The inequality originally derived by Bell is a statistical expression in three cross-correlated random observables each assumed to be a function of the same random variables. Two of the three observables occur at mutually exclusive settings on one side of a Bell apparatus but were defined by Bell to be alternative results of which only one could be observed. Both observables depended on the same hidden variable values. Thus, in Bells formulation, the correlations in the inequality were not all actually observable since the mutually exclusive alternatives occurred for the same values of the hidden variables. This situation is greatly simplified when it is discovered that the same inequality in correlations that Bell derived from a purely theoretical construction holds as an identity for correlation estimates using any three finite data sets, corresponding to Bell's three assumed observables. However, the data sets may now be local, nonlocal, random, deterministic, or nonsensical. The Bell inequality is an expression in algebra identically satisfied by data with values of + or -1. This implies that for laboratory data, when used consistently with Bell's derivation, there can be no Bell inequality violation. The same conclusion easily follows in the four variable case. The logical result is that there is no Bell's theorem as ordinarily understood.

The more general derivation of the inequality given above leads to an experimental procedure to obtain data for all the correlations, even for one involving the unobservable alternative that served in Bell's original formulation. When two mutually exclusive settings on one side of a Bell-experimental apparatus are employed in different experimental runs, the same inequality in correlations is now applicable.

There appear to be very few ways that three data observations on two particles destroyed by observation can be acquired. While measurements at alternate mutually exclusive settings for deterministic variables are routinely obtained in laboratories, the case of alternate random variables and associated probabilities requires that an experiment be repeated many times at each setting of interest. The novelty in the Bell experiment case is that outputs at each of the two alternate settings are correlated with each other through their correlation with a common output at a third variable setting. The three variables' correlations are now consistent with the resultant inequality in correlations.

Historically hidden in the statistical derivation of the inequality, the identity-inequality beginning with correlation estimates implies a corresponding inequality in the probabilities that produce them. In the quantum mechanical case that is the subject of the Bell theorem, quantum mechanical probabilities produce correlations that satisfy the Bell inequality while the probabilities satisfy the Wigner inequality that follows from it. The conclusion is made foolproof by the fact, as stated, that the Bell inequalities in either three variables (the original form), or four, are easily proven to be algebraic identities in the number of variables used to compute them.

The fact that the Bell inequalities in both three and four variables are algebraic identities has been unrecognized by experimentalists who have sought to test the inequalities experimentally. However, given the lack of recognition of the underlying basis of the inequalities in cross-correlation, correlations of independently observed data pairs have been inserted into them. The resulting correlations that must be different when consistent with inequality derivation and the quantum mechanical experiment, now all have the same functional form so as to result in inequality violation.

Only for a wide sense stationary process would the procedure used yield correlations that satisfied the inequalities, but they would then have a different functional form.

Finally, the Bell correlation is shown to be computable using an independent computer simulation without assuming physical variable characteristics commonly thought necessary to obtain cosine correlations. The observables use two functions for their definition and two qualitatively different random processes in the overall construction, the second reminiscent of that occurring in physical Bell sources. It is doubtful that this process is adequately represented by Bell's probability notation. This example also shows that entanglement is not a unique condition for the production of cosine correlations.

From the above, the Bell theorem as usually understood does not exist since the inequality is identically satisfied under the conditions of derivation. As a result. theoretically predicted correlations that violate the Bell inequality represent no three data sets that can exist.

**Acknowledgement**

The presentation of the material given above has been influenced by many discussions of the issues with Joe Foremen, personal communications with Karl Hess and Armen Gulian, and critiques of earlier papers on this topic by Michael Hall.

**Competing Interests**

The author did not receive support from any organization for the submitted work.

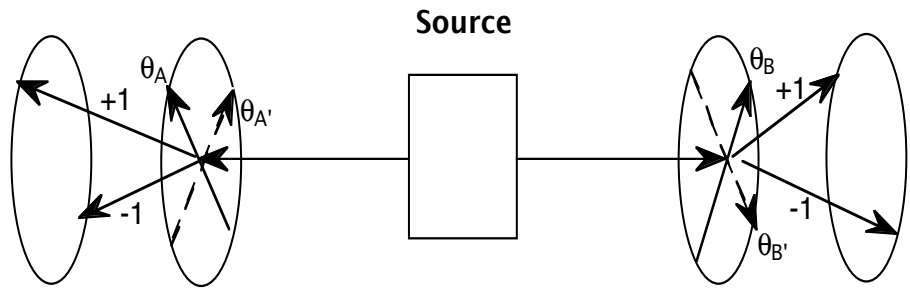

**Fig.** 1. Schematic of Bell experiment in which a source sends two particles (photons most often used) to two detectors having angular settings $\theta_A$ and $\theta_B$, (denoted as *a* and *b* in Bell's notation) and alternative settings $\theta_{A'}$ and $\theta_{B'}$. While one measurement operation on the *A*-side, e.g. at setting $\theta_A$, commutes with one on the *B*-side at $\theta_B$, any additional measurements at either $\theta_{A'}$ or $\theta_{B'}$ are non-commutative with prior measurements at $\theta_A$ and $\theta_B$, respectively. This figure was drawn by the author. and modified in notation for use in Ref. 4, as well as other papers.

## References
Fig. 1